\documentclass{PoS}

\title{Analytic continuation of the critical line in 2-color QCD at nonzero 
temperature and density}

\ShortTitle{Analytic continuation of the critical line in 2-color QCD at 
nonzero $T$ and $\mu$}

\author{Paolo Cea\\
        Dipartimento di Fisica, Universit\`a di Bari, 
        and INFN - Sezione di Bari, I-70126 Bari, Italy\\
        E-mail: \email{paolo.cea@ba.infn.it}}

\author{Leonardo Cosmai\\
        INFN - Sezione di Bari, I-70126 Bari, Italy\\
        E-mail: \email{leonardo.cosmai@ba.infn.it}}

\author{Massimo D'Elia\\
        Dipartimento di Fisica, Universit\`a di Genova, 
        and INFN - Sezione di Genova, \\I-16146 Genova, Italy\\
        E-mail: \email{Massimo.Delia@ge.infn.it}}

\author{\speaker{Alessandro Papa}\\
        Dipartimento di Fisica, Universit\`a della Calabria, 
        and INFN - Gruppo Collegato di Cosenza, \\I-87036 Rende, Italy\\
        E-mail: \email{papa@cs.infn.it}}

\abstract{We determine the pseudo-critical line in the temperature - chemical 
potential plane of 2-color QCD by direct Monte Carlo simulations and by 
analytic continuation from imaginary chemical potential.}

\FullConference{The XXV International Symposium on Lattice Field Theory\\
		 July 30-4 August 2007\\
		 Regensburg, Germany}

\begin{document}

\section{Introduction}

It is widely accepted that QCD undergoes a transition from the hadronic to 
the so-called ``quark-gluon plasma'' phase for large enough temperature, 
depending on the baryon chemical potential. Determining the 
shape and the nature of this transition (or pseudo-critical) line in the
temperature - chemical potential plane is of central interest in 
cosmology, in astrophysics and in the phenomenology of heavy ion collisions. 
The lattice approach, which is the natural tool to face a non-perturbative 
problem like this, is plagued, however, by the well-known ``sign problem'': 
for non-zero chemical potential the QCD fermion determinant becomes complex 
and the standard Monte Carlo importance sampling is unfeasible.

Several strategies have been invented to circumvent this problem (for a review, 
see~\cite{Phi05} and~\cite{Sch06}). Here, we concentrate on one of these 
approaches, the method of analytic 
continuation, first used in Refs.~\cite{Lom00} and \cite{HLP01}. The idea 
behind this method is very simple: numerical simulations are performed at 
{\it imaginary} chemical potential, $\mu=i\mu_I$, for which the fermion 
determinant is real, then Monte Carlo determinations are interpolated by a 
suitable function and finally this function is analytically continued to real 
values of $\mu$. 
This method is rather powerful since temperature and chemical potential 
can be varied independently and there is no limitation from increasing lattice 
size, as in methods based on reweighting. 
There is, however, an important drawback: the periodicity of the partition 
function and the presence of non-analyticities arising for imaginary values of 
the chemical potential~\cite{RW86} limit the region useful for numerical 
determinations to the strip $0\leq\mu_I/T<\pi/3$. 
This implies that the accuracy in the interpolation of the results at imaginary 
chemical potential has a strong impact on the extension of the domain of real 
$\mu$ values reachable after analytic continuation.

Although the method is designed to infer the behavior of an observable with 
the real chemical potential from the knowledge of its dependence on the
imaginary chemical potential, the idea to analytically continue the pseudo-critical 
line itself has been extensively applied~\cite{FP02,FP03,DL} (see Ref.~\cite{FP02} 
for a discussion on the reliability of this application of the method).

A control on the systematics of the method of analytic continuation and
possible insights for its improvement can be achieved by testing it in 
theories which do not suffer the sign problem, by direct comparison of the 
analytic continuation with Monte Carlo results obtained directly at real 
$\mu$~\cite{HLP01,GP04,Kim05,CCDP}. In Ref.~\cite{CCDP}, in particular,
a high-precision numerical analysis in SU(2) (or 2-color QCD) with $n_f=8$ 
degenerate staggered fermions has shown that, for temperatures above the
pseudo-critical one at zero chemical potential, the extrapolation to real $\mu$
improves considerably if ratio of polynomials are used instead of
simple polynomials in interpolating the behavior with imaginary $\mu$ of some 
test observables, this validating a proposal formulated in Ref.~\cite{Lom05}. 

In this work we extend the numerical analysis of Ref.~\cite{CCDP}
to the study of the analytic continuation of the pseudo-critical line. 
The strategy is the following:
\begin{itemize}
\item for several fixed values of the chemical potential, both real and imaginary,
we determine the (pseudo-)critical $\beta$'s by looking for peaks in the 
susceptibilities of a given observable;
\item we interpolate the determinations of the critical $\beta$'s at {\it imaginary}
chemical potential with an analytic function of $\mu$, to be then extrapolated
to real chemical potential;
\item we compare the extrapolation with the determinations of the critical 
$\beta$'s at real chemical potential.
\end{itemize}
The aim of this investigation is to verify the possibility to analytically
continue the pseudo-critical line and, in affirmative case, to optimize the choice
of the interpolating function to be used. For a better control of the systematics,
we have repeated the outlined strategy for three different observables
(chiral condensate, Polyakov loop, plaquette). 

\section{Theoretical background}

\begin{figure}
\includegraphics[width=0.49\textwidth]{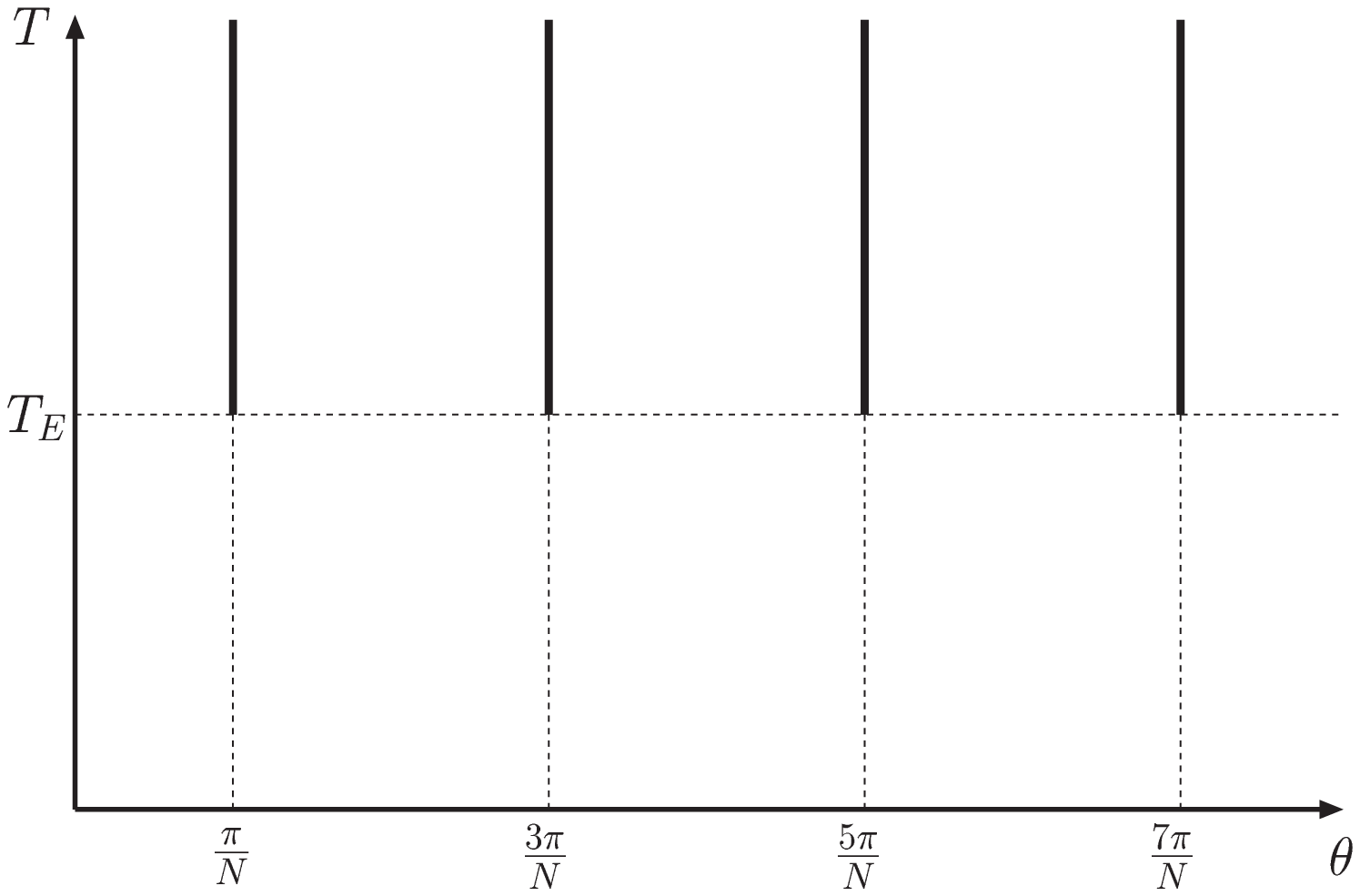}
\includegraphics[width=0.49\textwidth]{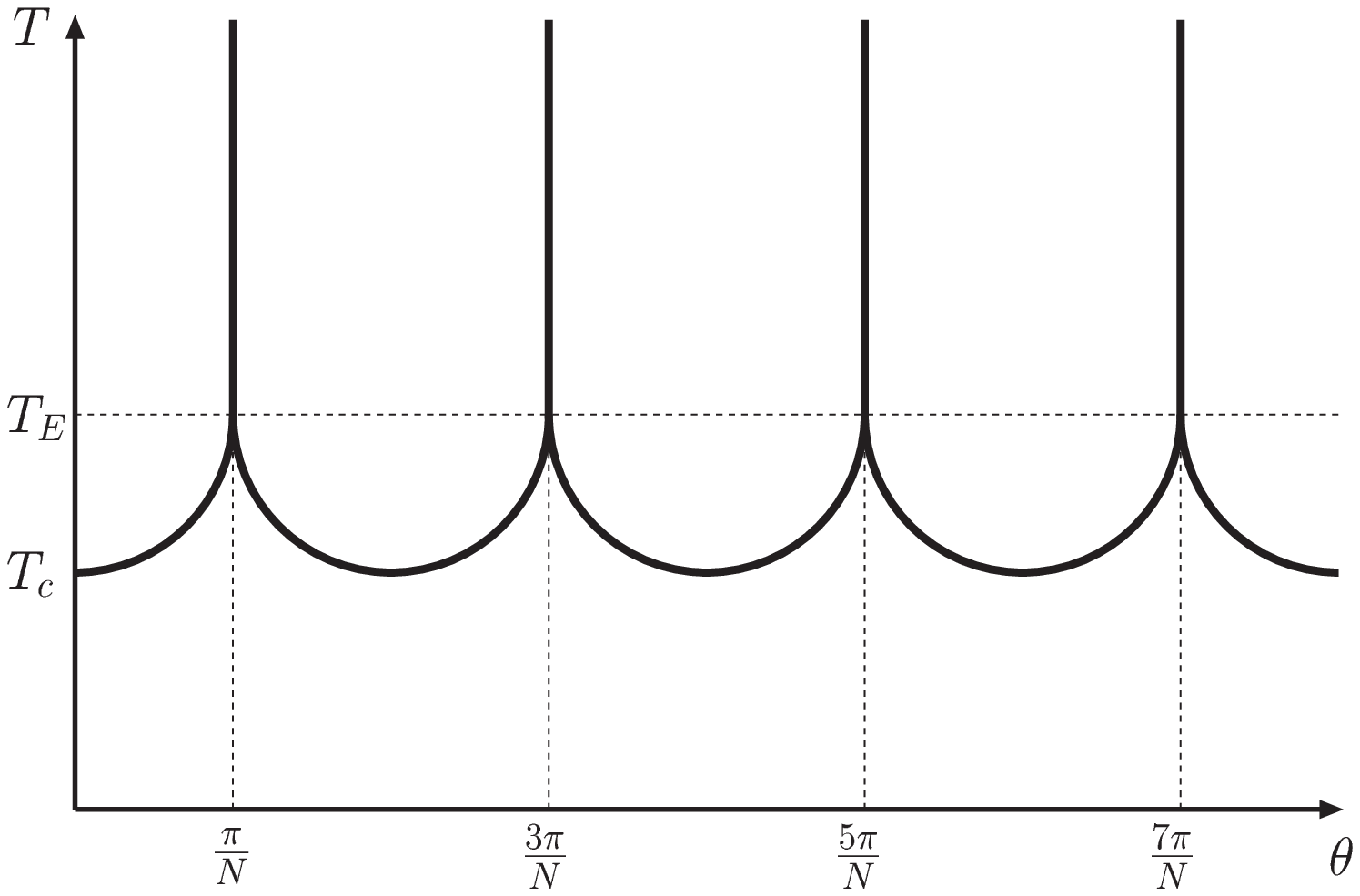}
\caption[]{(Left) Phase diagram in the $(T,\theta)$ plane according to 
Ref.~\cite{RW86}.
(Right) Tentative phase diagram in the $(T,\theta)$ plane after the inclusion 
of the chiral pseudo-critical lines.}
\label{fig1}
\end{figure}

\begin{figure}
\begin{center}
\includegraphics[width=0.65\textwidth]{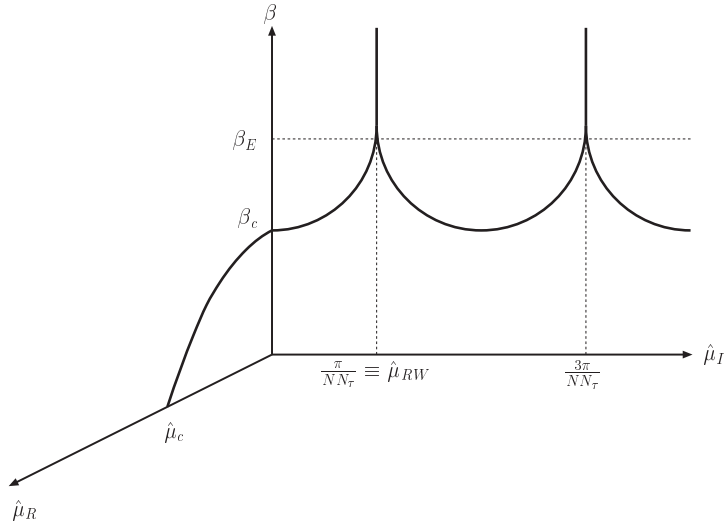}
\caption[]{Phase diagram in the $(\beta,\hat\mu_I,\hat\mu_R)$-space; $N$ is the number of colors, 
$N_\tau$ the extension of the lattice in the temporal direction.}
\label{fig2}
\end{center}
\end{figure}

Long ago Roberge and Weiss have shown~\cite{RW86} that the partition function of 
any SU($N$) gauge theory with non-zero temperature and imaginary chemical potential, 
$\mu=i\mu_I$, is periodic in $\theta\equiv\mu_I/T$ with period $2\pi/N$ and 
that the free energy $F$ is a regular function of
$\theta$ for $T < T_E$, while it is discontinuous at $\theta=2\pi(k+1/2)/N$, $k=0,1,2,\ldots$,
for $T > T_E$, where $T_E$ is a characteristic temperature, depending on the theory. The resulting
phase diagram in the $(T,\theta)$-plane is given in Fig.~\ref{fig1} (left), where the
vertical lines represent first order transition lines. This structure is compatible with the 
$\mu\to -\mu$ symmetry, related with CP invariance, and with the Roberge-Weiss
periodicity. The $\mu_I$-dependence of any observable is completely determined if this
observable is known in the strip $0\leq \theta < \pi/N$.  These predictions have been confirmed 
numerically in several cases, studying the behaviour of quantities like the Polyakov loop and 
the chiral condensate~\cite{FP02,DL,GP04}. 

A phase diagram like that in Fig.~\ref{fig1} (left) would imply the absence of any transition along 
the $T$ axis in the physical regime of zero chemical potential for any value of $N$, of $n_f$
and of the quark masses, which cannot be true. Therefore, it is necessary to admit that the phase
diagram in the $(T,\theta)$-plane is more complicated than in Fig.~\ref{fig1} (left). 
The simplest possibility is given in Fig.~\ref{fig1} (right), where the added lines generally
represent transitions which can be first order, second order or crossover. The temperature
$T_c$ is the pseudo-critical one for the transition at zero chemical potential.
In Fig.~\ref{fig2} the phase diagram of Fig.~\ref{fig1} (right) has been redrawn the 
$(\beta,\hat \mu_I)$-plane, where $\beta=2 N/g^2$, $\hat \mu_I=a \mu_I$ is the imaginary chemical 
potential in lattice units and it has been used the fact that $T=1/(a N_\tau)$, with $N_\tau$ the 
temporal extension of the lattice. In Fig.~\ref{fig2} also the $\mu_R$-axis has been included,
with a sketch of the continuation of the pseudo-critical line on the $(\beta,\hat \mu_R)$-plane.

\begin{figure}[tb]
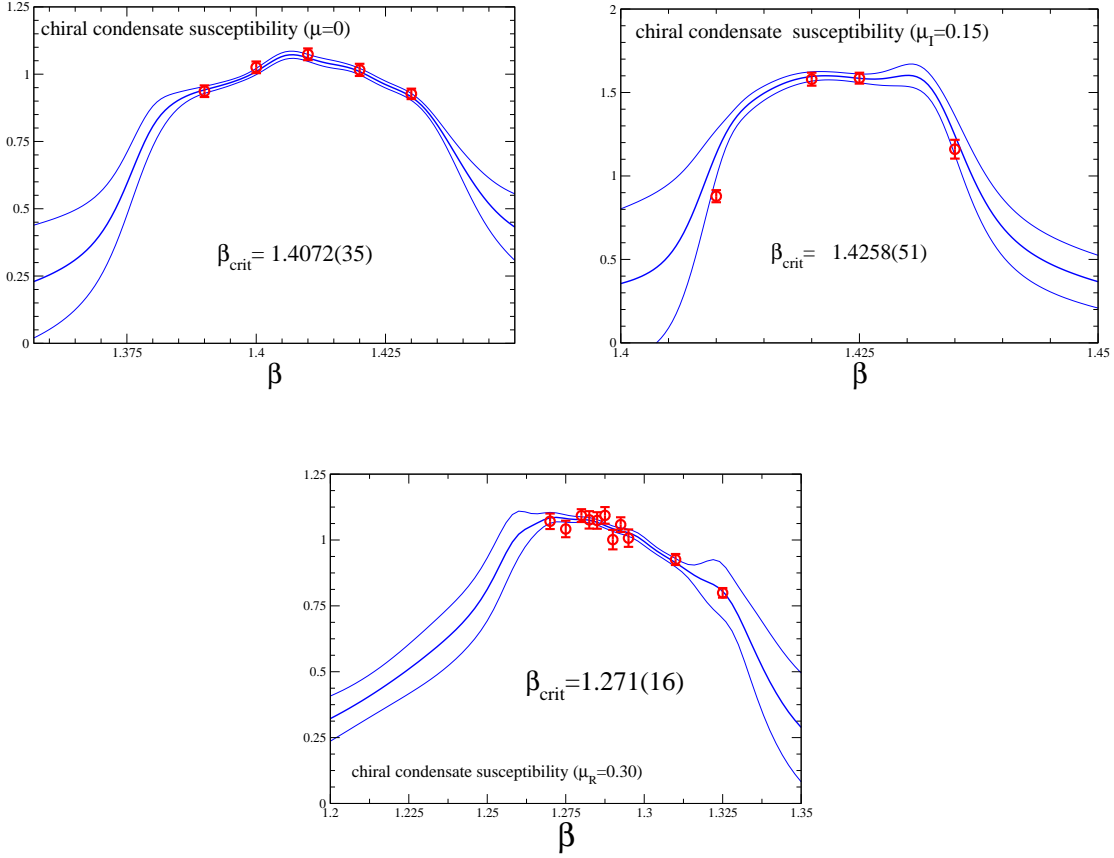

\begin{center}
\vspace{0.6cm}
\includegraphics[width=0.45\textwidth]{chiralsuscep_mu_0.eps}\hspace{1cm}
\includegraphics[width=0.45\textwidth]{chiralsuscep_muimm_015.eps}\vspace{1cm}
\includegraphics[width=0.45\textwidth]{chiralsuscep_mureal_030.eps}
\caption[]{Susceptibility of the chiral condensate at $\hat\mu$=0., $\hat\mu=0.15 i$ and 
$\hat\mu=0.30$. The blue solid lines represent the result (ant the related error) of the 
multi-histogram reweighting.}
\label{fig:susc_psibpsi}
\end{center}
\end{figure}

\section{Numerical results}

We performed numerical simulations on a $16^3\times 4$ lattice of the SU(2) gauge theory with 
$n_f=8$ degenerate staggered fermions having mass $am=0.07$, by means of a Hybrid Monte Carlo 
algorithm with $dt=0.01$. Simulations have been performed on the APEmille crate in Bari and on the 
computer facilities at the INFN apeNEXT Computing Center in Rome.

The observables we have considered are the chiral condensate, the Polyakov loop and the plaquette;
for each of them we have looked for the peak in the susceptibility for varying $\beta$, at some
fixed values of the chemical potential, both real and imaginary. For each observable and for each 
fixed value of $\hat\mu$, we have taken a few data points for different $\beta$ values 
(statistics $\sim$ 20000) and smoothed out the susceptibility by the multi-histogram extension 
of the Ferrenberg-Swendsen reweighting method~\cite{Ferrenberg:1988yz}. The uncertainty on
the position of the peaks has been evaluated by the bootstrap method.

\begin{table*}[tb]
\setlength{\tabcolsep}{1pc}
\centering
\caption[]{$\beta_{\mbox{\scriptsize crit}}$ as determined from the peak of the 
susceptibilities of chiral condensate, Polyakov loop and plaquette.} 
\vspace{0.1cm}
\begin{tabular}{llll}
\hline
\hline
$\hat\mu$ & chiral condensate & Polyakov loop & plaquette  \\  
\hline
0.30$\,i$ & 1.5097(33)        & 1.5023(60)    & 1.5048(62) \\
0.20$\,i$ & 1.4548(42)        & 1.4355(47)    & 1.4476(94) \\
0.15$\,i$ & 1.4258(51)        & 1.441(23)     & 1.418(10)  \\
0.10$\,i$ & 1.4228(55)        & 1.4089(63)    & 1.415(10)  \\
0.        & 1.4072(35)        & 1.394(17)     & 1.4060(45) \\
0.20      & 1.3551(91)        & 1.352(13)     & 1.356(12)  \\
0.30      & 1.271(16)         & 1.267(26)     & 1.286(15)  \\
\hline
\hline
\end{tabular}
\label{tab:peaks}
\end{table*}

\begin{table*}[tb]
\setlength{\tabcolsep}{1pc}
\centering
\caption[]{Parameters of the fit with a polynomial $A+B\hat\mu^2$ of the data at imaginary 
chemical potential; the last column gives the extrapolation of the pseudo-critical line at
the imaginary chemical potential corresponding to the first RW transition line.}
\vspace{0.1cm}
\begin{tabular}{lllll}
\hline
\hline
observable & $A$ & $B$ & $\chi^2$/d.o.f. & $\beta_{\mbox{\scriptsize crit}}(\hat\mu_{RW})$ \\  
\hline
chiral condensate & 1.4071(27) & $-$1.140(50) & 0.91 & 1.5829(81) \\
Polyakov loop     & 1.3931(55) & $-$1.18(10)  & 0.85 & 1.575(17)  \\
plaquette         & 1.4042(39) & $-$1.104(83) & 0.48 & 1.574(13)  \\
\hline
\hline
\end{tabular}
\label{tab:fit}
\end{table*}

\begin{figure}[tb]
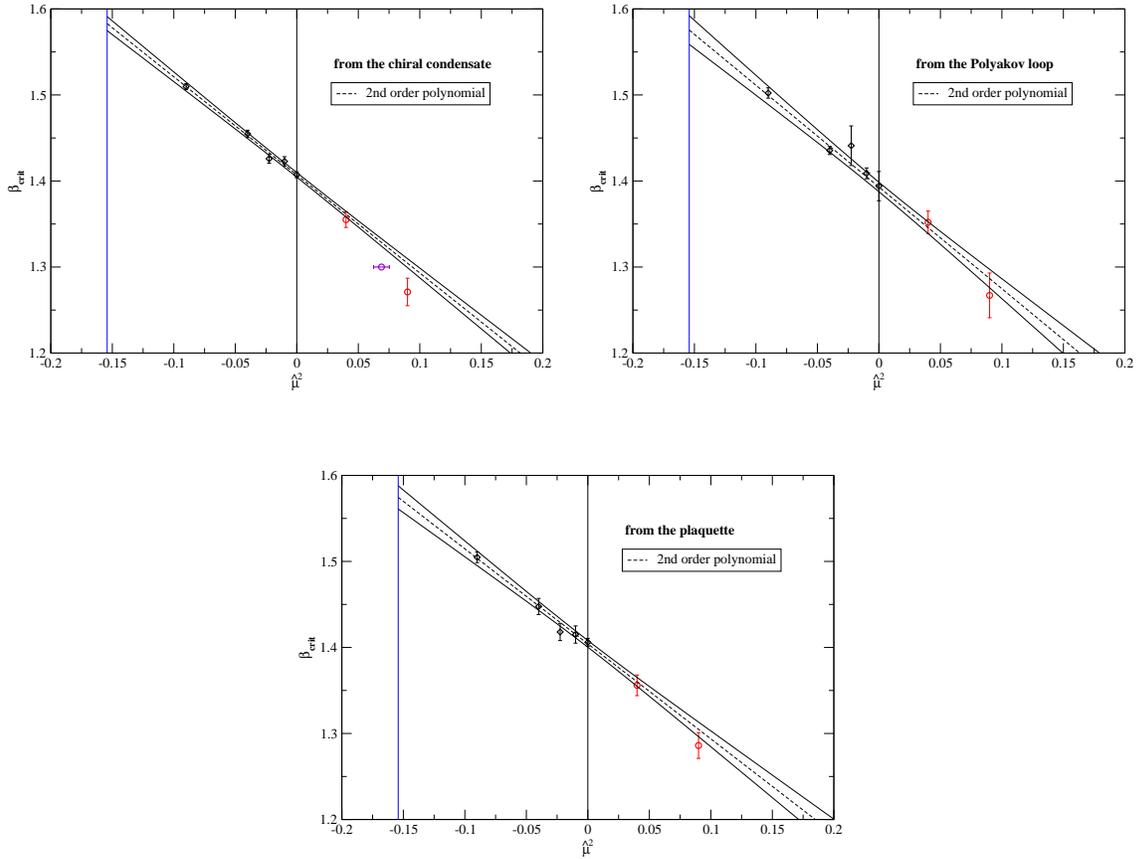

\begin{center}
\includegraphics[width=0.48\textwidth]{psibpsi.eps} \quad
\includegraphics[width=0.48\textwidth]{poly.eps}

\vspace{1cm}
\includegraphics[width=0.48\textwidth]{plaq.eps}
\caption[]{$\beta_{\mbox{\scriptsize crit}}$ {\it vs.} $\hat\mu^2$ determined from the 
susceptibility of chiral condensate, Polyakov loop and plaquette. 
The dashed line is the result of a fit with a polynomial 
$A+B\hat\mu^2$ of the data at imaginary chemical potential ($\hat\mu^2<0$); the black solid lines 
enclose the boundary of the uncertainty. The blue vertical line gives the position of the 
first RW transition line. The point in magenta comes from a determination of the peak of the 
susceptibility at $\beta=1.30$, taken from Ref.~\cite{CCDP}.}
\label{fig:crit}
\end{center}
\end{figure}

In Table~\ref{tab:peaks} we summarize our preliminary results for the critical $\beta$
at each value of the chemical potential we considered. In Fig.~\ref{fig:susc_psibpsi} we show 
for illustration purposes the susceptibility of the chiral condensate at $\hat\mu$=0.15$i$, 
$\hat\mu$=0. and $\hat\mu$=0.30. 

We have then looked for an interpolation of the data of $\beta_{\mbox{\scriptsize crit}}$ 
for {\it imaginary} and zero chemical potential (i.e. for the first 5 entries in 
Table~\ref{tab:peaks}), and have repeated this procedure for each of the three observables 
considered. 
In all cases we have found that the optimal interpolating function is a polynomial 
of the form $A+B\hat\mu^2$. If different functions are used, such as larger order polynomials or
ratio of polynomials, the fit puts to values compatible with zero all parameters except two of them, 
so to reduce the interpolating function to a first order polynomial in $\hat\mu^2$.
The fit results are summarized in Table~\ref{tab:fit}. We can see that the resulting parameters
have a tiny dependence on the observable considered; moreover, the extrapolation
of the pseudo-critical line at the $\hat\mu$ value corresponding to the first Roberge-Weiss
transition line, $\hat\mu_{RW}=i \pi/8$, is in good agreement with an independent determination
of the endpoint $\beta_E$~\cite{Cor07}.
This is a confirmation of the structure of the phase diagram as sketched in Fig.~\ref{fig1} (right).

The most important point of the present analysis is to test whether the extrapolation of 
the pseudo-critical line to real $\hat\mu$ agrees or not with the two direct determinations of
$\beta_{\mbox{\scriptsize crit}}$ available so far at real $\hat\mu$, i.e. $\hat\mu$=0.20 and 0.30.
Such comparison is presented in Figs.~\ref{fig:crit} for each of the observables considered. 
There is an overall substantial agreement; in the case of the chiral condensate there might be
a deviation at $\hat\mu$=0.30, which calls for a refinement of the numerical analysis.  

\section{Conclusions and outlook}

We have presented preliminary results aimed at studying the possibility
of the analytic continuation of the pseudo-critical line from imaginary to real 
chemical potential in 2-color QCD. This theory is exempt of the sign problem
and, therefore, makes possible to compare analytic continuations from
imaginary to real chemical potential with direct determinations at real chemical
potential.

By determining the position of the peaks in the susceptibilities of three observables
(chiral condensate, Polyakov loop, plaquette) for varying the temperature at fixed
imaginary chemical potential, we have interpolated the pseudo-critical line in the 
temperature - imaginary chemical potential plane. It turns out that the best interpolation
for $\beta_{\mbox{\scriptsize crit}}(\mu^2)$ is a first order polynomial in $\mu^2$. The 
extrapolation to real chemical potential of this curve generally agrees with the direct 
determinations at real chemical available so far. A larger statistics and an extension of
the data set could reveal possible deviations.

\end{document}